\documentclass[twocolumn,aps,prl,showpacs,preprintnumbers,amsmath,amssymb,superscriptaddress]{revtex4-1}
\usepackage[latin9]{inputenc}
\usepackage[switch]{lineno} 
\usepackage{color}
\usepackage{array}
\usepackage{verbatim}
\usepackage{prettyref}
\usepackage{multirow}
\usepackage{amsmath}
\usepackage{graphicx}
\usepackage{esint}
\usepackage{epstopdf}
\usepackage{upgreek}
\usepackage{tabularx} 
\usepackage{tikz}
\usetikzlibrary{arrows}
\usepackage[unicode=true,
 bookmarks=false,
 breaklinks=false,pdfborder={0 0 1},colorlinks=true] 
 {hyperref}
\hypersetup{
 pdftex,pdfstartview=FitV,linkcolor=blue,citecolor=blue,urlcolor=blue,plainpages=true}
\usepackage{booktabs}
\usepackage{feynmp} 	
\usepackage{ifpdf}
\ifpdf
\fi

\makeatletter

\providecommand{\tabularnewline}{\\}

\usepackage{dcolumn}
\usepackage{bm}
\usepackage{subfigure}

\makeatother

\begin{document}

\title{Exciton condensate in bilayer transition metal dichalcogenides: strong coupling regime}

\author{Bishwajit Debnath}

\affiliation{Department of Electrical and Computer Engineering, University of California, Riverside, CA 92521, USA}

\author{Yafis Barlas}

\affiliation{Department of Physics and Astronomy, University of California, Riverside, CA 92521, USA}

\author{Darshana Wickramaratne}

\affiliation{Materials Department, University of California, Santa Barbara, CA 93106, USA}

\author{Mahesh R. Neupane}

\affiliation{Electronics Technology Branch, Sensors and Electron Devices Directorate, U. S. Army Research Laboratory, Adelphi, MD 20783, USA}

\author{Roger K. Lake}

\affiliation{Department of Electrical and Computer Engineering, University of California, Riverside, CA 92521, USA}
\email{rlake@ee.ucr.edu}

\begin{abstract}
Exciton condensation in an electron-hole bilayer system of monolayer transition metal dichalcogenides
is analyzed at three different levels of theory to account for screening
and quasiparticle renormalization.
The large effective masses of the transition metal dichalcogenides place them in a 
strong coupling regime.
In this regime, mean field (MF) theory with either an unscreened or screened
interlayer interaction predicts a room temperature condensate.
Interlayer and intralayer interactions
renormalize the quasiparticle dispersion, and this
effect is included in a $GW$ approximation. 
The renormalization reverses the trends predicted from the unscreened or screened MF
theories.
In the strong coupling regime, intralayer interactions have a large impact on the
magnitude of the order parameter and its functional dependencies on effective mass and
carrier density. 

\end{abstract}
\maketitle

\section{Introduction}

\begin{figure}[t]
\vspace{0.55cm}
\centering{}\includegraphics[width=3.1in]{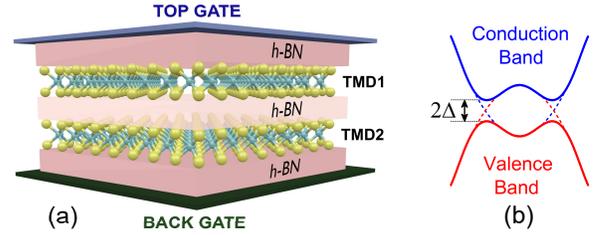} \caption{\label{fig:device} 
(a) Illustration of two monolayers of transition metal dichalcogenides separated
by a thin film of hexagonal boron nitride (h-BN). 
The Fermi levels
of the top and bottom monolayers are tuned 
to induce equal electron and hole carrier densities. 
(b) The conduction band of the electron layer and the valence band of the
hole layer overlap, and in the presence of a condensate, a gap
($2\Delta$) opens in the dispersion.}
\end{figure}

Electron-hole (e-h) bilayer systems, such as the one illustrated in Fig.~\ref{fig:device}(a),
are good candidates for observing exciton condensation~\cite{lozovik_1976_electron_hole_bilayer}.
The presence of an exciton condensate results in a gapped spectrum for the e-h bilayer system, 
as illustrated in Fig.~\ref{fig:device}(b).
Although there is evidence of exciton 
condensation in GaAs double quantum wells in the quantum Hall regime~\cite{1994_Butov_Condense_GaAs,PhysRevLett.84.5808,PhysRevLett.93.036801,PhysRevLett.93.036802}, the zero-field exciton condensate remains elusive. 
Recently, focus has returned to engineering a bilayer exciton condensate in the absence of
a magnetic field in
two-dimensional crystals such as graphene 
and transition metal dichalcogenides 
\cite{Hongki_RT_Superfluid_BLG,2008_Zhang,2008_Kharitonov,2013_superfluidity_bilayer_graphene,2013_Sarma_interlayer_exciton_superfluidity,Fischetti_BLExciton_PRB14,2D_indirect_exciton_McDonald_2015,MacDonald_Spatially_Indirect_PRB17}. 

Graphene appears to be an attractive candidate for the realization of bilayer
exciton condensates due to its perfect particle-hole nesting~\cite{Hongki_RT_Superfluid_BLG,2008_Zhang}.
Mean field calculations with the bare Coulomb interaction predict high transition temperatures ($\sim 300$ K)~\cite{Hongki_RT_Superfluid_BLG}.  
However, screening effects in graphene are of the order of the
Fermi wavevector ($k_{F}$). 
As a result, static screening reduces the transition 
temperatures significantly~\cite{2008_Kharitonov,2012_gorbachev_nature,Fischetti_BLExciton_PRB14}.
The predicted transition temperatures in the e-h graphene bilayer systems
range from 1 mK -- 100 K 
\cite{Hongki_RT_Superfluid_BLG,2008_Zhang,Lozovik2008,2008_Kharitonov,2009_Lozovik,2010_Lozovik,2011_Mink,Fischetti_BLExciton_PRB14}, 
depending on the level of the theory.
A study which includes dynamical effects on the screened interactions estimates 
a transition temperature $T_{c} \sim 4$ K \cite{Sodemann_BLG_TI_2012}.
Another study taking into account the screening resulting from proximity gates
found transition temperatures in the 1 mK--1 K range \cite{Fischetti_BLExciton_PRB14}.
Replacing each monolayer of graphene with a bilayer of graphene 
has been suggested for increasing the transition temperature
\cite{2014_double_few_layer_graphene}. 

The strength of the exciton condensate is proportional to 
the coupling strength $\lambda$, which is
the ratio of the interaction energy to the band energy.
This ratio is the fine structure constant in graphene given by 
$\lambda=e^{2}/[\kappa \hbar v_{F}] 
\sim 2.2/\kappa$ \cite{Hongki_RT_Superfluid_BLG,2012_Lozovik_Condensation_two_layer_graphene,Sodemann_BLG_TI_2012}, 
where $\kappa$ is the dielectric constant of the barrier material and $v_{F}$
is the Fermi velocity. 
Graphene's fine structure constant is density independent and typically 
$\lambda \lesssim 1$, which is a good approximation for weak coupling theories.
However for parabolic bands, such as those in bilayer 
graphene and transition metal dichalcogenides (TMDs), $\lambda$ is density dependent.
In this case, 
$\lambda \eqsim 6 g m_{+}/(\kappa k_{F})$, 
where $m_{+}$ is the reduced electron-hole mass of the e-h bilayer system,
$g$ is the degeneracy, and $k_F \propto \sqrt{n_{2D}/g}$ is the Fermi 
momemtum that depends on electron density $n_{2D}$.
In bilayer graphene, 
the low effective mass gives $\lambda = 0.2 \sim 1.1$, 
so that weak coupling theories also apply.

TMDs have larger effective masses and typically larger 
values of $\lambda = 2.2 \sim 10.4$, depending on the carrier density of $10^{11}\sim 10^{12}$ cm$^{-2}$.
Larger masses result in larger excitonic binding energies
that would appear more suitable for higher exciton gaps and transition temperatures.
Mean field calculations using the unscreened Coulomb interactions do predict
room temperature condensation, 
and they also predict higher condensation temperature
for higher carrier densities ($n_{2D}$).
However, for higher carrier densities, screening effects should be considered.
In graphene bilayers, 
screening incorporated within a random phase approximation (RPA)
reduces the interlayer coherence, as one would expect.
For TMD bilayers, which lie in the strong coupling regime, RPA screening
has little effect on the interlayer coherence.
Screening not only affects the {\em interlayer} interaction, but it also
affects the {\em intralayer} interaction within the same monolayer.
The intralayer interaction renormalizes the effective mass and
the corresponding $\lambda$.
We formulate an intermediate/strong coupling theory by incorporating 
both the interlayer and the intralayer RPA screened interaction
into a self-energy correction that renormalizes both the effective masses 
and the excitonic gaps.
The inclusion of the self-energy renormalization 
reverses the trends predicted from the unscreened and screened MF theories.
In the weak coupling limit, 
the intermediate/strong coupling theory converges
to the MF theory with an unscreened interaction.

The remainder of the paper is organized as follows. 
Section II describes the effective model for TMDs used in this paper. 
Section III discusses the standard mean field treatment of the model
Hamiltonian for the bilayer TMD system with an unscreened interaction. 
In section IV, we include RPA screening and a self-energy renormalization
in a $GW$ approximation and compare the predictions of the different levels of theory.
Section V summarizes and concludes.

\begin{table*}[t]
\caption[TMDC Material Parameter]{\footnotesize{
TMD material parameters obtained using density functional theory (HSE-SOC) \cite{Darshana_thesis}. 
$m_\alpha$ is the effective mass at the valence band edge ($K_{v}$) and
the conduction band edge ($K_{c}$), in the units of free electron mass $m_0$.
$\kappa$ is the relative dielectric constant. 
$n_{2D}$ and $k_F$ are the maximum allowed electron density and Fermi wavevector for one-type of spin determined
by the conduction spin-splitting energy $\Delta_{c}$.
}}
\centering\label{table:material_parameter}
\bgroup
\def\arraystretch{1.2} 	
\begin{tabularx}{0.8\textwidth}{c @{\extracolsep{\fill}} cccccccc}
\hline 
\hline 
\multirow{2}{*}{Material} & \multicolumn{3}{c}{Effective Mass ($m_{\alpha}$)} & \multicolumn{2}{c}{Band Splitting} & $\kappa$ & $n_{2D}$ & $k_{F}$\tabularnewline
\cline{2-9}
 & \multicolumn{1}{c}{Direction} & $K_{v}$ & $K_{c}$ & $\Delta_{v}(meV)$ & $\Delta_{c}(meV)$ &  & $(\times10^{12} \, cm^{-2})$ & $(nm^{-1})$\tabularnewline
\hline 
\multirow{2}{*}{MoS$_{2}$} & Longitudinal & 0.485 & 0.407 & \multirow{2}{*}{188.6} & \multirow{2}{*}{9.9} & \multirow{2}{*}{3.43} & \multirow{2}{*}{0.4} & \multirow{2}{*}{0.1585}\tabularnewline
 & Transverse & 0.480 & 0.404 &  &  &  &  & \tabularnewline
\hline 
\multirow{2}{*}{MoSe$_{2}$} & Longitudinal & 0.503 & 0.435 & \multirow{2}{*}{254.8} & \multirow{2}{*}{36.9} & \multirow{2}{*}{4.74} & \multirow{2}{*}{1.7} & \multirow{2}{*}{0.3268}\tabularnewline
 & Transverse & 0.503 & 0.436 &  &  &  &  & \tabularnewline
\hline 
\multirow{2}{*}{MoTe$_{2}$} & Longitudinal & 0.576 & 0.501 & \multirow{2}{*}{317.4} & \multirow{2}{*}{43.7} & \multirow{2}{*}{5.76} & \multirow{2}{*}{2.3} & \multirow{2}{*}{0.3801}\tabularnewline
 & Transverse & 0.565 & 0.500 &  &  &  &  & \tabularnewline
\hline 
\multirow{2}{*}{WS$_{2}$} & Longitudinal & 0.304 & 0.331 & \multirow{2}{*}{528.7} & \multirow{2}{*}{12.0} & \multirow{2}{*}{4.13} & \multirow{2}{*}{0.4} & \multirow{2}{*}{0.1585}\tabularnewline
 & Transverse & 0.305 & 0.332 &  &  &  &  & \tabularnewline
\hline 
\multirow{2}{*}{WSe$_{2}$} & Longitudinal & 0.303 & 0.358 & \multirow{2}{*}{606.4} & \multirow{2}{*}{7.80} & \multirow{2}{*}{4.63} & \multirow{2}{*}{0.3} & \multirow{2}{*}{0.1373}\tabularnewline
 & Transverse & 0.303 & 0.359 &  &  &  &  & \tabularnewline
\hline 
\multicolumn{1}{c}{} & \multicolumn{1}{c}{} & \multicolumn{1}{c}{} & \multicolumn{1}{c}{} & \multicolumn{1}{c}{} & \multicolumn{1}{c}{} & \multicolumn{1}{c}{} & \multicolumn{1}{c}{} & \multicolumn{1}{c}{}\tabularnewline
\end{tabularx}
\egroup
\end{table*}

\section{Effective model for e-h TMDs bilayers}

We consider several TMD electron-hole bilayers separated by an 
insulating h-BN spacer layer, as illustrated in Fig ~\ref{fig:device}(a). 
Separation of the electron and hole layers by a barrier
reduces the overlap of their respective wavefunctions
which reduces the interlayer tunneling and recombination. 
The Fermi level lies in the conduction band of the top monolayer
and in the valence band of the bottom monolayer.

The two layers of the bilayer system can consist of the same TMDs (homo-bilayer)
or different TMDs (hetero-bilayer). 
To achieve high critical temperatures for exciton 
condensation particle-hole nesting is beneficial, (\textit{i.e.},  $|m_{e}| = |m_{h}|$).
The electron and hole masses in TMDs are similar but not equal,
therefore, we consider different homo- and hetero-layer 
TMD combinations. 

\begin{figure}[t]
\centering{}\centering{\includegraphics[width=3.35in]{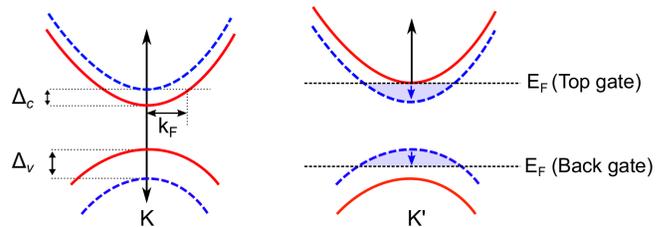}}
\caption{\label{fig:band-alignment} 
Spin composition at $\mathbf{K}$ and $\mathbf{K}'$ of monolayer 
MoX$_2$ TMDs \cite{Falko_kp_2DMat15}. 
Up- and down-spin bands are denoted by solid-red
and dash-blue lines, respectively. 
Spin-orbit coupling causes
spin splitting of the conduction band ($\Delta_{c}$) and the
valence band ($\Delta_{v}$). 
}
\end{figure}

Table \ref{table:material_parameter} 
shows the spin-resolved band parameters, the effective masses
and maximum 2D carrier density for several monolayer TMDs, calculated
using spin-resolved density functional theory \cite{VASP}.
All the \textit{ab initio} calculations, including the geometric relaxation and 
the electronic band properties, 
are performed at the hybrid
Heyd-Scuseria-Ernzerhof (HSE) level of theory \cite{HSE1} with spin-orbit coupling (SOC).
Calculation details are described in Appendix A of \cite{Darshana_thesis}.
From the effective masses in Table \ref{table:material_parameter}, we identify 
several TMD bilayer combinations with partial electron-hole
nesting (\textit{i.e.}, $|m_{e}| \sim |m_{h}|$ ). 
All of the n-type layers are chosen from the MoX$_2$ materials with spin
splitting illustrated in Fig. \ref{fig:band-alignment} \cite{2011_SOC_in_MX2,Falko_kp_2DMat15}.
The spin splitting of the conduction band $\Delta_c$ sets the maximum
Fermi level for each calculation.
Within this limit, each band of each K-valley is spin polarized. 

Treating the electron and hole dispersions as parabolic, the model Hamiltonian
for the structure is $\mathcal{H}=\mathcal{H}_{0}+\mathcal{H}_{e-e}$, 
\begin{equation}
\mathcal{H}=\sum_{{\bf k}\sigma\alpha}\epsilon_{k,\sigma}^{\alpha}c_{{\bf k}\sigma\alpha}^{\dagger}c_{{\bf k}\sigma\alpha}+\frac{1}{2S}\sum_{{\bf q}\alpha\beta}v_{q}^{\alpha\beta}\rho_{\alpha}({\bf q})\rho_{\beta}(-{\bf q}),\label{Ham}
\end{equation}
where $c_{{\bf k},\sigma,e}^{\dagger}\,\,(c_{{\bf k},\sigma,h}^{\dagger})$
denote the electron (hole) creation operators,
$\sigma$ denotes the spin and valley quantum numbers for the electron/hole,
${\bf k}=(k_{x},k_{y})$ is the in-plane two-dimensional
momentum with $k=\sqrt{k_{x}^{2}+k_{y}^{2}}$, 
$S$ is the area of the bilayer, 
$\alpha(\beta) \in \{e,h\}$
are the electron/hole layer indices, 
$\epsilon_{k,\sigma}^{\alpha=e}=\hbar^2 (k^{2}-k_{F}^{2})/(2m_{e,\sigma})$,
$\epsilon_{k,\sigma}^{\alpha=h}=-\hbar^2 (k^{2}-k_{F}^{2})/(2m_{h,\sigma})$,
$k_{F}$ is the Fermi momentum, and 
$m_{e(h),\sigma}$ denotes the spin and valley dependent effective
masses for the electron (hole).
Time reversal symmetry dictates that $m_{\alpha, \sigma} = m_{\alpha, -\sigma}$.
In Eq. (\ref{Ham}), 
$\rho_{\alpha}({\bf q})=\sum_{{\bf k}\sigma}c_{{\bf k}+{\bf q}\sigma\alpha}^{\dagger}c_{{\bf k}\sigma\alpha}$
is the total electron density for the $\alpha^{th}$ layer, $V_{ee}=V_{hh}=2\pi e^{2}/(\kappa q)$
is the Fourier transform of the intralayer interaction, and $V_{eh}=V_{he}=-V_{ee}e^{-qd}$
is the Fourier transform of the interlayer interaction, where $\kappa$
is barrier dielectric constant, $d$ is the thickness of the h-BN insulating
spacer, and $q$ is the momentum transfer, $q=|{\bf {k}-{\bf {k}'|}}$.

\begin{figure*}[t]
\vspace{-0.5cm}
\centering \includegraphics[height=2.2in]{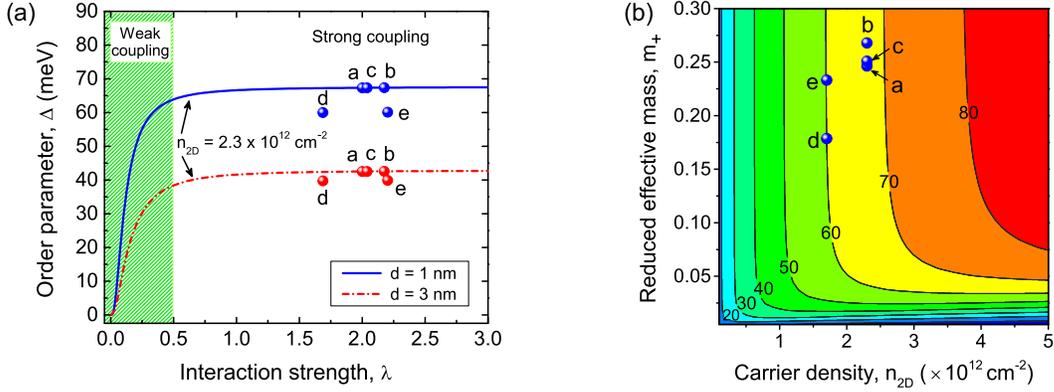}
\vspace{-0.2cm}\caption{\label{fig:BareGap} 
(a) The order parameter $\Delta$
as a function of the interaction strength $\lambda=ge^{2}m_{+}/(\pi\kappa\hbar^{2}k_{F})$
for $d = 1.0$ nm (solid line), and $d = 3.0$ nm (dashed line) with $n_{2D}=2.3\times10^{12}\,$cm$^{-2}$. 
The spheres give the order parameter for five possible e/h bilayers:
a) MoTe$_{2}$/MoS$_{2}$, b) MoTe$_{2}$/MoTe$_{2}$, c) MoTe$_{2}$/MoSe$_{2}$,
d) MoSe$_{2}$/WSe$_{2}$, e) MoSe$_{2}$/MoSe$_{2}$. 
In calculating $\Delta$ for specific combinations, respective effective masses and
the maximum allowed 2D carrier density of the electron layer are used. 
The interaction strength $\lambda$ for the TMD bilayers lies in the strong coupling regime.
(b) Color contour plot of $\Delta$ as a function of $m_{+}$ and $n_{2D}$ with $d = 1.0$ nm. 
The value in meV of each contour is labeled.
The positions of the 5 bilayers of (a) are shown.
}
\end{figure*}

\section{Mean Field Theory}
\label{sec:mean_field}
Mean field decomposition of Eq. \eqref{Ham}, gives an effective BCS-like Hamiltonian.
The Green's function for the MF effective Hamiltonian can be expressed as,
\begin{equation}
\widehat{G}_0\,(k,\omega) =
\frac{(\omega - \zeta_k) \mathcal{\hat{I}} + \xi_{k}\hat{\tau}_{3} + \Delta_{k}\hat{\tau}_{1}}
{(\omega - \zeta_k)^{2} - E_{k}^{2} + i\eta}
\label{eq:G_k_omega},
\end{equation}
where $\hat{\tau}_{i}$ is a Pauli matrix representing the layer pseudospin
in the indices $\alpha$ and $\beta$; 
$\zeta_k = \hbar^2 k^{2} / (4m_{-,\sigma})$, 
$m_{-,\sigma}^{-1}=(m_{e,\sigma}^{-1}-m_{h,\sigma}^{-1})$,
$\xi_{k}=\hbar^2 (k^{2}-k_{F}^{2})/(4m_{+,\sigma})$, 
$m_{+,\sigma}^{-1}=(m_{e,\sigma}^{-1}+m_{h,\sigma}^{-1})$,
$E_{k}=\sqrt{\xi_{k}^{2}+\Delta^{2}}$, 
and $\Delta_k$ is the order parameter.
When $\Delta\rightarrow0,$ the Green function in Eq. \eqref{eq:G_k_omega}
reduces to the Green's function of the normal state. 
The value of the 
order parameter $\Delta$ is evaluated self-consistently, 
\begin{equation}
\Delta_{{\bf {k}}}=-\frac{1}{2}\sum_{{\bf {k}'}}V_{eh}(|{\bf k}-{\bf {k}'}|)\frac{\Delta_{{\bf {k}'}}}{E_{{\bf {k}'}}}.
\end{equation}

In general, the order parameter can have a complicated dependence on momentum,
but here we assume a translationally invariant order parameter
$\Delta$. 
We evaluate the normalized order parameter $\bar{\Delta}=\Delta/\epsilon_{F}$, as a function of the interaction strength $\lambda$
and the interlayer separation $d$,
%
\begin{equation}
1= \lambda \intop_{-\pi/2}^{\pi/2} d\phi \intop_{0}^{2\cos\phi}d\bar{q} \,\,\,
\frac{ v_D(q) }{ \sqrt{ {\bar{\xi}_{k-q}}^2+\bar{\Delta}^{2} } },
\label{eq:delta}
\end{equation}
where 
$\bar{\xi}_{k-q}=\xi_{k-q}/\epsilon_F$, 
$v_D(q) = e^{-k_{F}\bar{q}d}$, 
$\lambda=g e^{2}m_{+}/(\pi \kappa\hbar^{2}k_{F})$, 
$\kappa$ is the dielectric constant of the h-BN barrier (3.9), 
$g$ is flavor multiplicity (two-fold for the valley degeneracy), 
$\phi$ is the angle between $\mathbf{k}$ and $\mathbf{q}$, 
and $\bar{q}=q/k_{F}$ when $k_{F}=\sqrt{4\pi n_{2D}/g}$. 
Note the appearance of the interaction parameter $\lambda$,
which captures the strength of the interlayer
coherence.
Eq. \eqref{eq:delta} is evaluated self-consistently at $k=k_{F}$.
Henceforth, we restrict our attention to the case where the electron and
hole densities are identical, $n_{e}=n_{h}=n_{2D}$.
We refer to this approach as unscreened mean field (MF) and will denote it as MF.

Figure \ref{fig:BareGap}(a) shows the dependence of the order parameter $\Delta$ as
a function of the coupling parameter $\lambda$ at a carrier density
of $n_{2D}=2.3\times10^{12}\,\text{cm}^{-2}$. 
Eq. (\ref{eq:delta})
predicts that room temperature condensation
is possible for $\lambda \gtrsim 0.2$. 
Due to the exponential dependence of $d$ in Eq. \eqref{eq:delta}, 
decreasing the interlayer separation from 3 nm to 1 nm increases 
the order parameter by almost a factor of two. 
Figure \ref{fig:BareGap}(a) also shows the order parameter $\Delta$ for five possible 
TMD bilayer structures (blue/red spheres):
a) MoTe$_{2}$/MoS$_{2}$, b) MoTe$_{2}$/MoTe$_{2}$, c) MoTe$_{2}$/MoSe$_{2}$,
d) MoSe$_{2}$/WSe$_{2}$, e) MoSe$_{2}$/MoSe$_{2}$. 
The order parameters for these combinations are calculated using 
the masses and maximum carrier densities of the n-type layer as listed
in Table \ref{table:material_parameter}.  
Due to the higher effective masses
and lower carrier densities,
the values of $\lambda$ for these bilayer combinations
are in the strong coupling 
regime $(\lambda \sim 2)$. 
Figure \ref{fig:BareGap}(b) shows the order parameter 
$\Delta$ in the phase space of the reduced effective mass ($m_+$) and the electron density ($n_{2D}$).
The positions of the 5 bilayer systems are also shown.
As anticipated, the unscreened mean field theory indicates that 
exciton condensation is favorable for higher 2D carrier densities and
larger effective masses.
The unscreened mean field calculations are generally
valid for weak coupling regimes ($\lambda \sim 0.5$).
Considering that the TMD hetero-structures fall in the strong coupling
regime ($\lambda \sim 2$), the theory of exciton condensates in TMDs 
must be enhanced to include screening and renormalization effects.
In the next section, we formulate a strong coupling theory that includes
screening of the Coulomb interaction, as well as the effect of quasiparticle 
renormalization.

\section{Intermediate/Strong coupling theory}
 
In this section, we first include RPA screening and then 
self-energy renormalization in a $GW$ approximation. 
Results from the different levels of theory are compared.

\subsection{Screened interlayer and intralayer interaction}
\begin{figure}[t]
\centering
\vspace{0.2cm}
\includegraphics[width=3.2in]{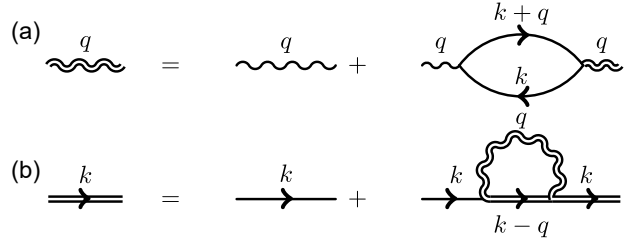}
\vspace{-0.2cm}
\caption{(a) 
Screened interaction in the RPA approximation. 
The Green's function used in the polarization bubble depends on the level of theory. 
(b)  Dyson equation for the Green function in a $GW$ approximation that includes both interlayer and intralayer screening.
}
\label{fig:Feynman}
\end{figure}

Screening is treated in the random phase approximation 
as illustrated in Fig. \ref{fig:Feynman}(a).
At this level of theory, 
the solid lines in the polarization diagram
represent the Green function of Eq. (\ref{eq:G_k_omega}) which includes
the coherence term $\Delta$. 
$\Delta$ is calculated from Eq. \eqref{eq:delta} using the screened interaction
self-consistently with the polarization functions.

The polarization is a $2\times2$ matrix with diagonal terms $\Pi_S$ corresponding
to same-layer polarization, and off-diagonal terms $\Pi_D$ corresponding to
different-layer polarization.
When the top and bottom layers have the same carrier density, 
$\Pi_{S}$ and $\Pi_{D}$ 
can be decoupled into even and odd channels, 
defined as $\Pi_{\pm}=\Pi_{S}\pm\Pi_{D}$ where
\begin{align}
\Pi_{\pm}(q,\omega)  =  g\int\frac{d^{2}k}{(2\pi)^{2}} & \bigg(\frac{E_{k+q}+E_{k}}{E_{k+q}E_{k}}\times\nonumber \\
& \frac{E_{k+q}E_{k}\mp\Delta^{2}-\xi_{k+q}\xi_{k}}{\omega^2+i\eta-(E_{k}+E_{k+q})^{2}}\bigg).
\label{eq:pi_pm}
\end{align}
The particle-hole response functions depend on the order parameter $\Delta$
and also through the
gapped spectrum $E_{k} = \sqrt{\xi_{k}^2 + \Delta^2}$,
as seen explicitly in Eq. \ref{eq:pi_pm}. 
The response function 
is evaluated self-consistently with the order parameter. 
From this point onward, we neglect dynamical retardation
of the screened interaction and set the frequency $\omega=0$.

The even polarization function $\Pi_{+}$ captures the density response 
to the total charge density of the gapped spectrum. 
Since the total response of a gapped system to a uniform shift in 
the potential vanishes, $\Pi_{+}(q \to 0)=0$.
The odd channel polarization function $\Pi_{-}$ captures the 
response to a difference in the charge density of the two layers.
In the $q \to 0$ limit, the odd channel polarization function 
approaches the density of states, $\Pi_{-}(q\to0,\omega=0)
=-N(\epsilon_{F})$, independent of the gap $\Delta$.

The intralayer and interlayer density response
functions needed for the calculations are $\Pi_{S} = (\Pi_+ + \Pi_-)/2$ and $\Pi_{D} = (\Pi_+ - \Pi_-)/2$ 
given by
\begin{align}
\Pi_{S}(q) &= g\int\frac{d^{2}k}{(2\pi)^{2}}\bigg(1-\frac{\xi_{k+q}\xi_{k}}{E_{k+q}E_{k}}\bigg)\times \frac{-2E_{k}}{(E_{k}+E_{k+q})^{2}}, \label{eq:pi_s} \\
\Pi_{D}(q) &= g\int\frac{d^{2}k}{(2\pi)^{2}}\frac{2\Delta^2}{E_{k+q}\,(E_{k}+E_{k+q})^{2}}.\label{eq:pi_d}
\end{align}
The response functions are normalized to the 2D density of states 
as $\Pi_{S(D)} (q) = -N(\epsilon_{F}) \, \chi_{S(D)} (q)$, 
where $N(\epsilon_{F}) = gm_{+}/(2\pi\hbar^{2})$ is the density of states for
the parabolic bands and $\chi_{S(D)} (q)$ are the dimensionless polarization functions.

The interlayer screened interaction $V_{eh}^{sc}(q)$, within the RPA, can be 
expressed as $V_{eh}^{sc}(q)=2\pi e^{2}/(\kappa q)\cdot v_{D}^{sc}(\bar{q})$,
where 
\begin{equation}
v_{D}^{sc}(\bar{q})=\resizebox{0.8\hsize}{!}{$\frac{\bar{q}\,\,\bigr[v_{D}+ \tilde{\lambda} \, (v_{S}^{2}-v_{D}^{2}) \, \chi_{D}\bigr]}{1- 2\tilde{\lambda} (v_{S}\chi_{S}+v_{D}\chi_{D})+ \tilde{\lambda}^2 (v_{S}^{2}-v_{D}^{2})(\chi_{S}^{2}-\chi_{D}^{2})}.$}\label{eq:screened_VD}
\end{equation}
Here, we define $v_{S}=1/\bar{q}$, $v_{D}=e^{-k_{F}\bar{q}d}/\bar{q}$ 
and $\tilde{\lambda} = 2 \pi \lambda$. 
In the limit of an unscreened potential, 
$v_{D}^{sc}(\bar{q})$ reduces to $v_D(q) = e^{-k_{F}\bar{q}d}$ of Eq. \eqref{eq:delta}.

One can now include self-consistent screening in the calculation of the order parameter by
replacing the bare Coulomb potential $v_{D}(q)$ 
in Eq. \eqref{eq:delta} with the screened interlayer
interaction $v_{D}^{sc}(q)$,
and calculate $\Delta$ in Eq. \eqref{eq:delta}, $\Pi_S$ in Eq. \eqref{eq:pi_s}, 
$\Pi_D$ in Eq. \eqref{eq:pi_d}, and $v_{D}^{sc}(q)$ in Eq. \eqref{eq:screened_VD}
self-consistently. 
We refer to this approach as mean field with RPA screening (MF-RPA).

Electron-electron interactions not only result in screening,
but they also renormalize the quasiparticle dispersion.
The self-energy renormalization is affected by both the
interlayer and the intralayer interactions.
Similar to the screened interlayer interaction in
Eq. \eqref{eq:screened_VD}, the screened intralayer
interactions are $V_{ee}^{sc}(q)=V_{hh}^{sc}(q)=2\pi e^{2}/(\kappa q) \cdot v_{S}^{sc}(\bar{q})$,
where
\begin{equation}
v_{S}^{sc}(\bar{q})=\resizebox{0.8\hsize}{!}{$\frac{\bar{q}\,\, \bigr[v_{S} -\tilde{\lambda}\,(v_{S}^{2}-v_{D}^{2})\, \chi_{S}\bigr]}{1-2\tilde{\lambda}(v_{S}\chi_{S}+v_{D}\chi_{D})+\tilde{\lambda}^2(v_{S}^{2}-v_{D}^{2})(\chi_{S}^{2}-\chi_{D}^{2})}.$}
\label{eq:screened_Vs}
\end{equation}
This correctly reduces to the monolayer RPA interaction in the limit $d \to \infty$.

The order parameter is directly proportional to the interlayer screened potential $v_{D}^{sc}$ . 
The intralayer interaction $v_S^{sc}$ enters into the diagonal element of the self-energy
which renormalizes the quasiparticle dispersion ($\xi_{k}$)
and the interaction strength $\lambda$.
To understand these effects, we 
determine the self-energy of Fig. \ref{fig:Feynman}(b)
and use it to calculate the order parameter self-consistently.

\subsection{Self-energy correction to many-body interaction}
\begin{figure}[t]
\centering \vspace{0.2cm}
\includegraphics[width=2.8in]{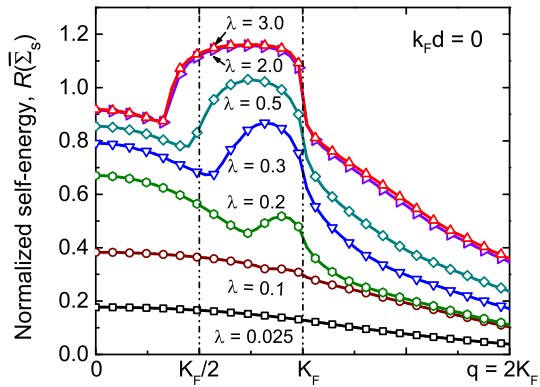}
\caption{Real part of the diagonal self-energy $\Sigma_S(q)$ normalized to $\epsilon_F$ for $k_{F}d=0$.}
\label{fig:sigmaBar_vs_q}
\end{figure}

The renormalization of both the quasiparticle dispersion and the interlayer interaction
are included within a $GW$ approximation.
The self-energy illustrated in Fig. \ref{fig:sigmaBar_vs_q} 
is calculated self-consistently with the Green's function.
The Green's functions used in the polarization diagram include the renormalized order parameter
but ignore the mass renormalization.
Only the real part of the self energy is used in the calculation
of the Green's function.
We refer to this approach as mean field with $GW$ renormalization (MF-GW).

Denoting the $2\times2$ self-energy matrix as $\widehat{\Sigma}_c$,
the full Green function matrix $\widehat{\mathcal{G}}(k,\omega)$
is given by
$\widehat{\mathcal{G}}^{-1}(k,\omega)=\widehat{G}^{-1}(k,\omega)-\widehat{\Sigma}_{c}\,(k,\omega)$,
where $\widehat{G}$ is the
bare Green function in Eq. \eqref{eq:G_k_omega}.
Hence, the full Green function is
\begin{equation}
\widehat{\mathcal{G}}^{-1}=\left[\begin{array}{cc}
\omega+i\eta-[\xi_{k}+\mathcal{R}(\Sigma_{S})] & -\Delta_0-\mathcal{R}(\Sigma_{D})\\
-\Delta_0-\mathcal{R}(\Sigma_{D}) & \omega+i\eta+[\xi_{k}+\mathcal{R}(\Sigma_{S})]
\end{array}\right]\label{eq:complete_G_k_omega},
\end{equation}
where $\Delta_0$ is the gap function in the absence of the self-energy correction,
and $\mathcal{R}$ denotes the real part.
It is clear from Eq. (\ref{eq:complete_G_k_omega}) that
the diagonal element $\Sigma_{S}$ renormalizes the quasiparticle 
dispersion as $\xi_{k}\rightarrow\xi_{k}+\mathcal{R}(\Sigma_{S})$,
and the off-diagonal element $\Sigma_{D}$ renormalizes the gap function 
as $\Delta_0 \to \Delta_0 + \mathcal{R}(\Sigma_D)$.

We calculate the diagonal self-energy as
\begin{equation}
\Sigma_{S}\,(k,\omega-\Omega)=i\int\frac{d\Omega}{2\pi}\int\frac{d^{2}q}{\left(2\pi\right)^{2}}v_{S}^{sc}(q)\,\widehat{\mathcal{G}}_{S}(k-q,\omega-\Omega)\label{eq:Sigma_Diagonal} ,
\end{equation}
where $\widehat{\mathcal{G}}_{S}$ is the diagonal part of the Green's
function in Eq. \eqref{eq:complete_G_k_omega}. 
We take the complex path integral
over $\Omega$ in Eq. \eqref{eq:Sigma_Diagonal} and calculate the
normalized diagonal self-energy in the static limit ($\omega\rightarrow0$),
\begin{eqnarray}
\mathcal{R}(\overline{\Sigma}_{S}(\bar{k})) = -\left(\frac{\lambda}{\pi}\right)\intop_{0}^{2\pi}d\phi\intop_{0}^{2} & d\bar{q} &\,v_{S}^{sc}(\bar{q})\Theta(k_{F}^{2}-|\textbf{k}-\textbf{q}|^{2})\nonumber \\
 & \times & \frac{-|\bar{\xi}^{R}_{\bar{k}-\bar{q}}|}{\sqrt{\left[\bar{\xi}^{R}_{\bar{k}-\bar{q}}\right]^{2}+\bar{\Delta}^{2}}},
\label{eq:Sigma_SC}
\end{eqnarray}
where 
$\bar{\xi}^{R}_{\bar{k}-\bar{q}}=\bar{k}^{2}-2\bar{q}\bar{k}\cos\phi+\bar{q}^{2} - 1 +
\mathcal{R}\{\overline{\Sigma}_{S}(\bar{k}-\bar{q})\}$
takes into account the renormalization of the quasiparticle dispersion.
$\Theta$ is the unit step function, and \textbf{$\overline{\Sigma}_{S}=\Sigma_{S}/\epsilon_{F}$}.
Since $\overline{\Sigma}_{S}(\bar{k})$ in Eq. \eqref{eq:Sigma_SC}
requires the evaluation of $\overline{\Sigma}_{S}(\bar{k}-\bar{q})$, we
use analytical continuation properties, 
\emph{i.e.}, $\mathcal{R}(\overline{\Sigma}_{S}(\bar{k}))=\mathcal{R}(\overline{\Sigma}_{S}(-\bar{k}))$. 
A separate calculation of the off-diagonal self-energy $\Sigma_{D}$ is avoided
by self-consistently absorbing it in the definition of $\bar{\Delta}$, 
\begin{equation}
1=\lambda\intop_{-\pi/2}^{\pi/2}d\phi\intop_{0}^{2\cos\phi}d\bar{q}\,\,\,\frac{\,v_{D}^{sc}(\bar{q})\,\,\,}{\sqrt{[\bar{\xi}^{R}_{1-\bar{q}}]^{2}+\bar{\Delta}^{2}}}\label{eq:gap-equation-SC}.
\end{equation}

The value of ${\Delta}$ determined from Eq. \eqref{eq:gap-equation-SC} is used 
self-consistently in determining the polarization functions $\Pi_S$ and $\Pi_D$ and thus
the screened interactions $v_{D}^{sc}$ and $v_{S}^{sc}$.
The dispersion represented by $\xi_k$ used in the calculation of the polarization functions
is the bare dispersion in the absence of $\Sigma_S$.
Thus, the Green function lines in the polarization bubble are partially self-consistent in that
they include the effect of the self-energy on the off-diagonal order parameter, but they do not
include the effect of mass renormalization.
Eqs. \eqref{eq:pi_s}, \eqref{eq:pi_d}, 
\eqref{eq:screened_VD}, 
\eqref{eq:screened_Vs}, 
\eqref{eq:Sigma_SC}, 
and \eqref{eq:gap-equation-SC} are the
set of self-consistent equations that are solved to obtain $\Delta$.

To understand the relative contribution of the self-energy correction, 
we plot the normalized $\mathcal{R}\,(\Sigma_S)$ 
in Fig. \ref{fig:sigmaBar_vs_q}
for different values of $\lambda$. 
In the weak coupling regime ($\lambda < 0.2$), 
the self-energy is only 20\% -- 60\% of the Fermi energy. 
However, at the onset of intermediate/strong coupling region ($\lambda \geq 0.5$), 
the self-energy becomes equal to or larger than the Fermi energy.
This illustrates the importance of the self-energy correction in the
strong coupling regime.

\subsection{Discussion}

\begin{figure}[t]
\centering \hspace{-0.25cm}
\includegraphics[height=2.1in]{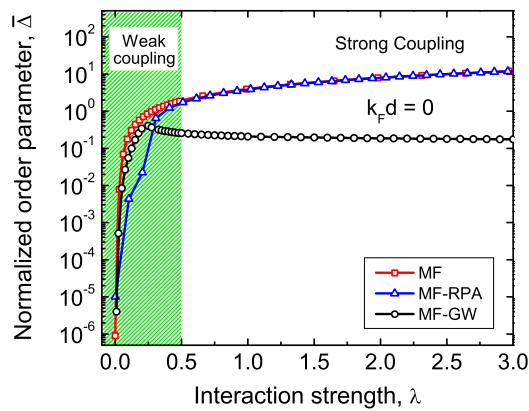}
\caption{Normalized order parameter as a function of the effective interaction strength
$\lambda$, obtained from MF, MF-RPA, and MF-GW theory for $k_{F}d=0$.}
\label{fig:screened_delta_vs_lambda}
\end{figure}

In this section, we discuss the MF-GW results and compare them
with the MF and MF-RPA predictions. 
Theoretically, the most favorable condition of condensation occurs at vanishing
interlayer distance, \emph{i.e.} $k_{F}d\rightarrow0$. 
Considering this optimum condition, 
Fig. \ref{fig:screened_delta_vs_lambda} summarizes
the three different levels of theory.
For the MF calculation 
the gap increases monotonically with interaction strength. 
In this case, moderate interlayer interaction
($\lambda>0.2$) leads to room temperature condensation.
The effect of RPA screening (MF-RPA) on the order parameter depends
on the relative strength of $\lambda$.
In the weak coupling regime ($\lambda\approx0.25$), screening
reduces the interlayer coherence.
In the intermediate/strong coupling regime, screening cannot
compete with the interlayer interaction and $\Delta$
follows the unscreened gap function. 
When both interlayer and intralayer screening 
are included as a self-energy correction (MF-GW), 
the interlayer coherence is strongly reduced for interaction strengths above 0.25.
As $\lambda \to 0$, the MF-GW theory and the MF theory coincide.
The reason is apparent from Fig. \ref{fig:sigmaBar_vs_q}, 
which shows that self-energy correction remains negligible up to $\lambda \sim 0.1$.

\begin{figure}[t]
\vspace{-0.45cm}\hspace{-0.3cm}\includegraphics[width=3.0in]{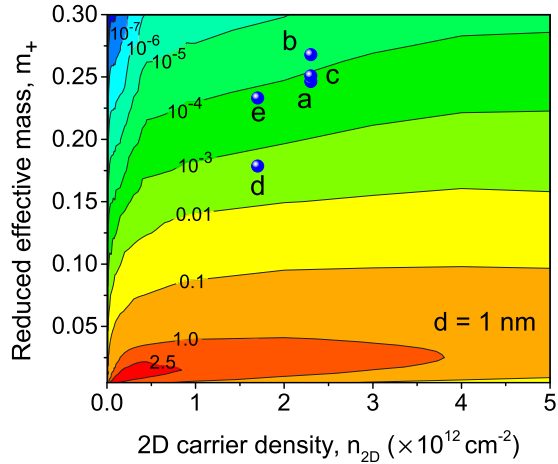} 
\caption{\label{fig:RPA_GW-gap} 
Color contour plot of $\Delta$ as a function of $m_{+}$ and $n_{2D}$ with $d = 1.0$ nm
calculated with MF-GW theory. 
The value in meV of each contour is labeled.
The positions of the 5 bilayers from Fig. \ref{fig:BareGap}(a) are shown.
}
\end{figure}

Figure \ref{fig:RPA_GW-gap} is
a color contour plot of $\Delta$ as a function of $m_+$ and $n_{2D}$ determined from the 
MF-GW theory.
The positions of the same bilayer structures from Fig. \ref{fig:BareGap} are shown.
A comparison of the $m_{+}-n_{2D}$ phase diagram in Fig. \ref{fig:RPA_GW-gap} with that of the MF result in 
Fig. \ref{fig:BareGap}(b) shows that MF-GW theory predicts trends that
are qualitatively different from the MF theory. 
For a reduced mass greater than 0.05, 
the order parameter of MF theory is nearly independent of the mass and is moderately dependent on the density,
changing by a factor of $\sim 3$ as the density increases an order of magnitude
from $5\times 10^{11}$ cm$^{-2}$ to $5\times 10^{12}$ cm$^{-2}$.
The order parameter of MF-GW theory has the same moderate dependence on the density, but it 
is exponentially dependent on the mass.
For a density of $2 \times 10^{12}$ cm$^{-2}$, the order parameter decreases 5 orders of magnitude as the mass
increases from 0.05 to 0.3.
Also, the functional dependence of the order parameter on the mass is qualitatively different.
In both theories, the order parameter rapidly increases as $m_+$ increases from zero.
In MF theory, the order parameter saturates and remains constant for $m_+ \gtrsim 0.1$.
In MF-GW theory, the order parameter peaks at $m_+ \sim 0.025$ and then exponentially decays as 
$m_+$ increases.
For MF theory, the conditions for maximum $\Delta$ occur at the upper right corresponding to high density and high mass.
For MF-GW theory, the conditions for maximum $\Delta$ occur at the lower left corresponding to 
low density and low mass.
The MF-GW theory exponentially reduces the magnitude of the order parameter for masses
corresponding to those of the 2D bilayers.
The heavy masses of the 2D materials which increase the order parameter in MF theory, decrease the order parameter
in MF-GW theory.

As shown in Fig. \ref{fig:screened_delta_vs_lambda}, 
{\em interlayer} screening calculated self-consistently in the presence of a condensate 
has little effect on the order parameter in the strong coupling limit.
Renormalization due to {\em intralayer} screening has a large effect.
We conclude that, in the strong coupling limit,
the {\em intralayer} interactions 
determine the overall trends of the order parameter.

\section{Conclusion}

Exciton condensation is analyzed
as a function of the coupling strength with a focus on the strong coupling regime, 
which is the regime of TMD bilayer electron-hole systems. 
Three different levels of theory are considered.
Starting from unscreened mean field theory,
RPA screening and self-energy renormalization in a $GW$ approximation are included.
A mean field calculation with an unscreened
Coulomb potential predicts a room temperature exciton condensate. 
The inclusion of RPA screening in the interlayer interaction
reduces the order parameter in the weak coupling regime,
but it has little effect in the strong
coupling regime, and a room temperature condensate is still predicted.
The inclusion of the effects of both the interlayer and intralayer interactions
through a self-energy correction to the quasiparticle dispersion and the order parameter
in a $GW$ approximation reverses the trends predicted from the MF and MF-RPA
theories.
The MF-GW theory favors low density and low mass for maximizing the magnitude of the
order parameter.
The heavy masses of the TMD materials that increase the order parameter in MF and MF-RPA
theories, reduce the order parameter in the MF-GW theory.
In the strong coupling regime, {\em intralayer} screening
has a large impact on the magnitude of the order parameter
and its functional dependencies on effective mass and carrier density.

\section*{Acknowledgements}
This work was supported in part by the National Science Foundation under Award NSF EFRI-1433395 and by
FAME, one of six centers of STARnet, a Semiconductor Research Corporation program sponsored by MARCO and DARPA.
This work used the Extreme Science and Engineering Discovery Environment (XSEDE), 
which is supported by National Science Foundation grant number ACI-1053575.


\end{document}